\documentclass[conference]{IEEEtran}
\IEEEoverridecommandlockouts
\usepackage{cite}
\usepackage{amsmath,amssymb,amsfonts}
\usepackage{algorithmic}
\usepackage{graphicx}
\usepackage{textcomp}
\usepackage{xcolor}
\def\BibTeX{{\rm B\kern-.05em{\sc i\kern-.025em b}\kern-.08em
    T\kern-.1667em\lower.7ex\hbox{E}\kern-.125emX}}

\graphicspath{{figures/}{./}} 

\usepackage{braket}

\usepackage[english]{babel}
\usepackage{float}
\usepackage{caption}
    
\newcommand{\half}{\frac{1}{2}}

\newcommand{\la}{\langle}
\newcommand{\ra}{\rangle}
\newcommand{\kb}{\ra\la}

\newcommand{\llp}{|l \ra\la l'|}
\newcommand{\lpl}{|l' \ra\la l|}
\newcommand{\llpPLUSlpl}{\llp + \lpl}

\newcommand{\sx}{\hat X}
\newcommand{\sy}{\hat Y}
\newcommand{\sz}{\hat Z}
\newcommand{\ident}{\hat I}

\newcommand{\Sx}{\hat S_x}

\newcommand{\Sz}{\hat S_z}
\newcommand{\Sint}{\hat S_z^{(i)} \hat S_z^{(j)}}

\newcommand{\SxSx}{\hat S_x^{(i)} \hat S_x^{(j)}}

\newcommand{\inttwo}{\hat{a}_i ^\dag \hat{a}_{i+1} + \textrm{h.c.}}

\newcommand{\ninj}{\hat n_i \hat n_j}
\newcommand{\qinj}{\hat q_i \hat n_j}
\newcommand{\qiqj}{\hat q_i \hat q_j}

\usepackage{mathtools}
\DeclarePairedDelimiter{\ceil}{\lceil}{\rceil}

\begin{document}

\title{On connectivity-dependent resource requirements for digital quantum simulation of $d$-level particles \\
}

\author{\IEEEauthorblockN{Nicolas P. D. Sawaya}
\IEEEauthorblockA{\textit{Intel Labs} \\
Santa Clara, CA 95054 \\
ORCID 0000-0001-8510-8480 \\
\textbf{nicolas.sawaya@intel.com}}
\and
\IEEEauthorblockN{Gian Giacomo Guerreschi}
\IEEEauthorblockA{\textit{Intel Labs} \\
Santa Clara, CA 95054, USA \\
ORCID 0000-0002-5579-451X}
\and
\IEEEauthorblockN{Adam Holmes}
\IEEEauthorblockA{\textit{Intel Labs} \\
Hillsboro, OR 97124, USA; and \\
\textit{University of Chicago} \\
Chicago, IL 60615, USA \\
ORCID 0000-0003-3770-1567}
}

\maketitle

\begin{abstract}
A primary objective of quantum computation is to efficiently simulate quantum physics. Scientifically and technologically important quantum Hamiltonians include those with spin-$s$, vibrational, photonic, and other bosonic degrees of freedom, \textit{i.e.} problems composed of or approximated by $d$-level particles (qudits). Recently, several methods for encoding these systems into a set of qubits have been introduced, where each encoding's efficiency was studied in terms of qubit and gate counts. Here, we build on previous results by including effects of hardware connectivity. To study the number of SWAP gates required to Trotterize commonly used quantum operators, we use both analytical arguments and automatic tools that optimize the schedule in multiple stages. We study the unary (or one-hot), Gray, standard binary, and block unary encodings, with three connectivities: linear array, ladder array, and square grid.  
Among other trends, we find that while the ladder array leads to substantial efficiencies over the linear array, the advantage of the square over the ladder array is less pronounced. 
These results are applicable in hardware co-design and in choosing efficient qudit encodings for a given set of near-term quantum hardware. Additionally, this work may be relevant to the scheduling of other quantum algorithms for which matrix exponentiation is a subroutine. 
\end{abstract}


\maketitle

\section{Introduction}
\label{sec:introduction}

Hamiltonian simulation---the simulation of quantum physics using a quantum computer---is likely to be a primary early application of quantum computation. Much theoretical work has been done on the general problem of Hamiltonian simulation as well as its application to problems in chemistry and materials science \cite{cao19,mcardle18_review}, condensed matter theory \cite{wecker15}, nuclear physics \cite{dumitrescu18}, and applications outside of physics \cite{gharibian15_qhamcompl}. Most work in this area has focused on fermionic and spin-$\half$ particles, although there is a large set of relevant problems of scientific interest involving ensembles of $d$-level systems (\textit{i.e.} qudits), including photonic \cite{sabin19,dipaolo19}, vibrational \cite{mahesh14,huh15,teplukhin18,sparrow18,sawaya19_vibr,mcardle19_vibr,wang19_vibr_supercond,magann20_control}, and spin-$s$ \cite{Levitt2008,lora_serrano_16} degrees of freedom.

In contrast to fermionic particles which require the use of Jordan-Wigner \cite{jw28} or related \cite{bk02,tranter15_bk,setia19_whitfield} transformations, here we are instead encoding $d$-level particles with bosonic commutation relations. This task consists primarily of mapping a series of local $d$-by-$d$ matrix operators to a set of qubits. For recent theoretical work on Hamiltonian simulation of photonic, vibrational, and bosonic degrees of freedom, the unary and standard binary (SB) encodings in second quantization have been considered \cite{somma2003arXiv,somma05,veis2016quantum,mcardle19_vibr,sawaya19_vibr,sabin19,chancellor19,geller19}, and a first quantization approach was also studied \cite{macridin18a,macridin18b}. A systematic study of $d$-level encoding approaches, using both previously used and novel encodings, was recently published \cite{sawaya19_dlev}, and the purpose of the current work is to build on these results.

In this work we study the effects of hardware connectivity on two-qubit operation counts, when simulating $d$-level systems on a digital quantum computer. Note that connectivity constraints have been considered previously for fermionic problems \cite{wecker15,kivlichan18_prl,berry18_improvedfermionic}. Most actively studied classes of quantum hardware \cite{krantz19_supercond, kloeffel13_qdots} (excluding ion trap systems \cite{bruzewicz19_ions}) do not inherently allow for operations between arbitrary pairs of qubits. Hence a series of SWAP gates must be performed to make relevant qubit pairs adjacent. Here we study which encodings are superior for a given set of problem parameters when hardware connectivity is taken into account, and determine the added utility of changing connectivity patterns. A portion of our results use a scheduler \cite{Haener2018,almudever2020} of quantum circuits that was previously reported \cite{guerreschi18_scheduling,guerreschi19_scheduler}.

\begin{figure}
    \centering
    \includegraphics[width=\linewidth]{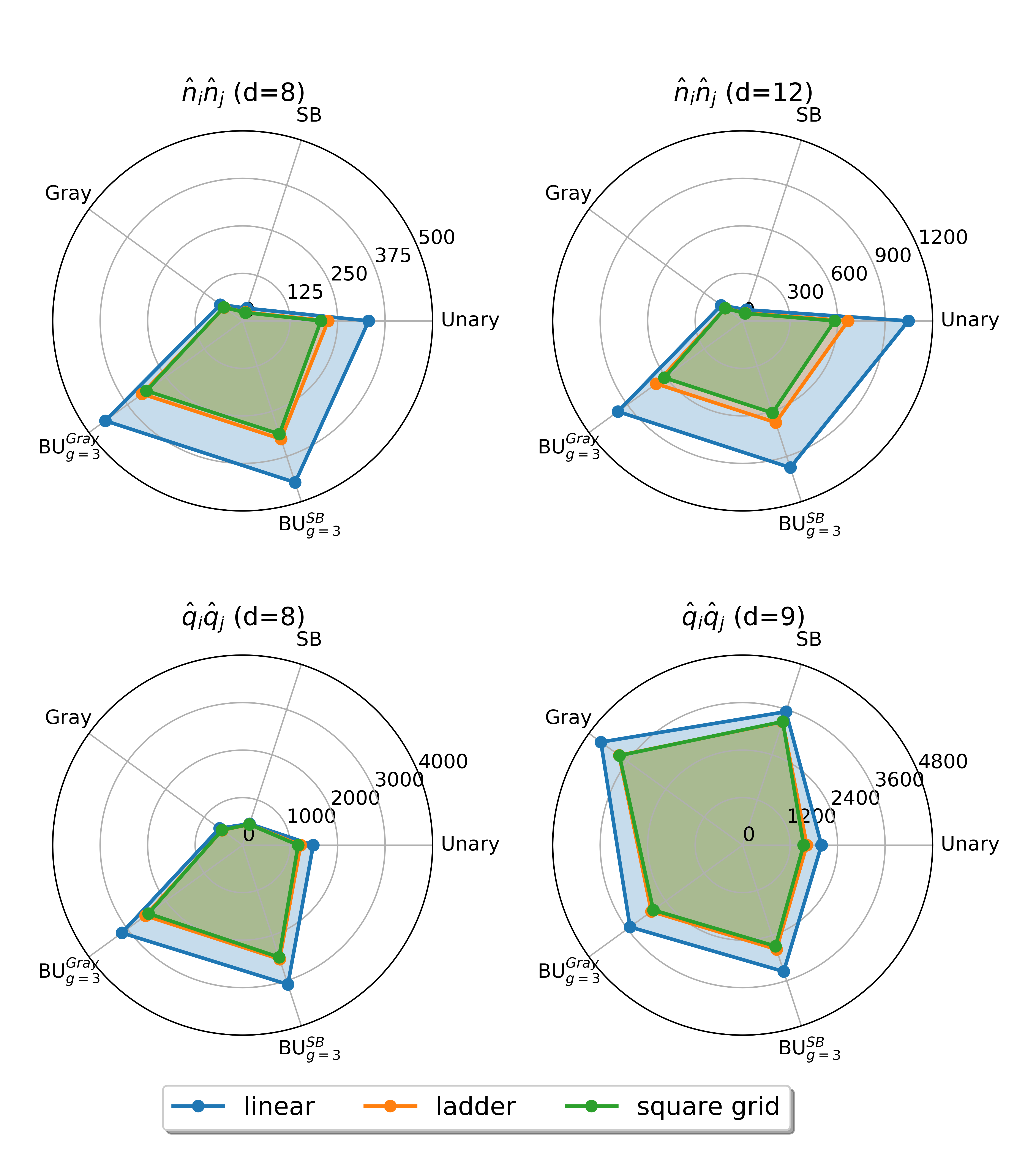}
    \caption{A representative selection of two-qubit gate counts (SWAP \& CNOT counts) for the two-particle operators $\ninj$ and $\qiqj$. Each direction on the radar chart represents a different encoding, with two values of $d$ shown for each operator. Each polygon represents a different connectivity. The blue (outermost), orange (middle), and green (innermost) polygons respectively represent linear, ladder, and square grid connectivity.}
    \label{fig:spider}
\end{figure}

A small subset of results are shown in Figure \ref{fig:spider}, for $d$-level two-particle operators $\ninj$ and $\qiqj$. The radius from the center of each radar chart equals the number of two-qubit gates and includes both the CNOTs needed by algorithm and the SWAPs required to overcome the limited connectivity. The blue (outermost) polygon of the radar chart represents linear hardware, with the inner polygons representing increased connectivity density. The plots demonstrate one of the main results of this work---namely, that the increase in connectivity from linear to ladder yields much more benefit than the subsequent transition to a 2D grid.

It is  useful to think of encoding choices in terms of a ``hardware budget''\cite{sawaya19_dlev}---the optimal encoding depends on both the coherence time and the number of available qubits. As some encodings may require more gates with fewer qubits while others require fewer gates but more qubits, the choice of encoding will often depend on the available quantum hardware. 

In Section \ref{sec:theory}, we summarize theory relevant to this work including encoding $d$-level particles to qubits. In Section \ref{sec:ub} we derive analytical upper bounds for the number of SWAP gates required to approximate a matrix exponential of local operators on hardware with linear connectivity. In Sections \ref{sec:methods} and \ref{sec:numerical-results} we respectively give our numerical methods and results. We end with discussion and outlook in section \ref{sec:concl}.

\section{Theory}
\label{sec:theory}

\subsection{Trotterization}
\label{sec:trotter}

Our goal is to implement the exponential of a Hamiltonian operator $\hat H$, a subroutine used both for dynamics and for eigenvalue estimation \cite{abrams99_qpe,kitaev97_qpe}. 
Begin with a Hamiltonian $\hat H$ acting on $N_p$ particles:
\begin{equation}
\hat H = \sum_i c_i \hat g_i
\end{equation}
where each term $c_i \hat g_i$ is a tensor product of $d$-level single-particle operators. Throughout this work, we assume that operators on different particles commute, as is the case for bosonic degrees of freedom.

In order to simulate the Hamiltonian on a qubit-based digital quantum computer, one must decompose the operator into a sum of Pauli strings such that
\begin{equation}
\hat H \mapsto \sum_j \hat h_j = \sum_j w_j \bigotimes_{k=1}^{N_q} \hat\sigma_{jk}
\end{equation}
where $N_q$ is the number of qubits and each $\hat\sigma_{jk} \in \{\sx_k,\sy_k,\sz_k,\ident_k\}$ is either a Pauli matrix or the identity on qubit $k$. A well-known example of this step is in the simulation of fermions, where the second quantized Hamiltonian is converted to a sum of Pauli strings using the Jordan-Wigner \cite{jw28}, Bravyi-Kitaev \cite{bk02,tranter15_bk}, or related transformations \cite{setia19_whitfield}. In simulation of $d$-level particles (qubits, truncated bosons, spin-$s$ particles), other approaches are instead required, as summarized in the next subsection. Several previous works have implemented mappings for qubit-based quantum simulation of bosons \cite{somma2003arXiv,somma05,veis2016quantum,macridin18a,macridin18b,mcardle19_vibr,sawaya19_vibr,sabin19,chancellor19,geller19,sawaya19_dlev}.

To approximate the exponential, one may implement \cite{whitfield10_elec,mikeike11} the Suzuki-Trotter formula \cite{suzuki76,Lloyd1996} 
\begin{equation}
\exp\left(-i \hat H \tau\right) \approx \left(\prod_j \exp(-i \hat h_j \tau/\eta) \right)^\eta
\end{equation}
which is exact in the limit of large $\eta$. Each Pauli string may be exponentiated using the well known ``CNOT staircase'' circuit \cite{mikeike11}, examples of which are shown in Figures \ref{fig:shuttle-cluster} and \ref{fig:unary_ordering}. 

\subsection{Encoding $d$-level systems}
\label{sec:enc}

We study the unary, standard binary, Gray, and block unary encodings. 
We refer to the Gray and standard binary encodings as \textit{compact} encodings. Examples for these encodings are shown in Tables \ref{tbl:encodingsbasic} and \ref{tbl:blockunary}. Because the Gray code's defining feature is a unity Hamming distance between consecutive integers \cite{roth06book}, it was shown (using all-to-all connectivity) to often require fewer entangling gates \cite{sawaya19_dlev} when Trotterizing common $d$-level operators.

Note that the qubit counts are not constant across encodings. Compact codes require $N_q=\ceil{\log_2 d}$ qubits, 
unary requires $N_q=d$, and block unary requires 
$N_q=\ceil{\frac{d}{g}} \ceil{\log_2(g+1)}$.

In the preceding expressions, $d$ is the number of levels in the particle, $g$ is the number of bits in one ``block,'' and 
$\ceil \cdot$ is the ceiling function. 
Because near-term hardware will be limited both in operations counts (due to decoherence times) and total qubits, the choice of encoding may be hardware-dependent. For instance, if one wants to simulate a Hamiltonian using a quantum computer with many qubits but shorter coherence time, an analysis might show that the unary code fits the hardware budget while a compact code does not.


\begin{table}[]
\centering
\begin{tabular}{lccc}
Decimal & Std. Binary & Gray Code & Unary \\
0  & \texttt{0000}  & \texttt{0000} & \texttt{000000001}  \\
1  & \texttt{0001}  & \texttt{0001} & \texttt{000000010}  \\
2  & \texttt{0010}  & \texttt{0011} & \texttt{000000100}  \\
3  & \texttt{0011}  & \texttt{0010} & \texttt{000001000}  \\
4  & \texttt{0100}  & \texttt{0110} & \texttt{000010000}  \\
5  & \texttt{0101}  & \texttt{0111} & \texttt{000100000}  \\
6  & \texttt{0110}  & \texttt{0101} & \texttt{001000000}  \\
7  & \texttt{0111}  & \texttt{0100} & \texttt{010000000}  \\
8  & \texttt{1000}  & \texttt{1100} & \texttt{100000000}  \\
\end{tabular}
\caption{SB, Gray, and unary encodings for decimal values 0 through 8.}
\label{tbl:encodingsbasic}
\end{table}

\begin{table}[]
\centering
\begin{tabular}{lcccc}
Decimal &    BU$_{g=3}^{SB}$ &   BU$_{g=3}^{Gray}$ &    \\
0  & \texttt{00 00 01}  & \texttt{00 00 01} & \\
1  & \texttt{00 00 10}  & \texttt{00 00 11} &  \\
2  & \texttt{00 00 11}  & \texttt{00 00 10} &  \\
3  & \texttt{00 01 00}  & \texttt{00 01 00} &   \\
4  & \texttt{00 10 00}  & \texttt{00 11 00} &   \\
5  & \texttt{00 11 00}  & \texttt{00 10 00} &   \\
6  & \texttt{01 00 00}  & \texttt{01 00 00} &   \\
7  & \texttt{10 00 00}  & \texttt{11 00 00} &  \\
8  & \texttt{11 00 00}  & \texttt{10 00 00} &   \\
\end{tabular}
\caption{Block unary encodings for decimal values 0 through 8.}
\label{tbl:blockunary}
\end{table}


Given a matrix element $| l \ra\la l' |$, one first converts the integers $l$ and $l'$ to bitstrings denoted $\mathcal R^{\textrm{enc}}(l;d)$ and $\mathcal R^{\textrm{enc}}(l';d)$ for some encoding. A relevant property of every non-compact encoding is that, to determine whether the system is in state $\ket{l}$, one may inspect only a subset of the qubits. Specifically, the bitmask subset $C(l) \equiv C^{\textrm{enc}}(l;d)$ determines which qubits must be included in the mapping of $\ket{l}$. For any integer $l$, the bitmask subset of a compact (SB or Gray) encoding includes all bits in the encoding, while that of a unary encoding includes just one bit. One may then write down the mapping
\begin{equation}
\label{eq:mapping}
\ket{l}\bra{l'} \mapsto \bigotimes_{i \in C(l) \cup C(l')} | x_i \ra\la x'_i |_i .
\end{equation}
%


For each factor in the right hand side of eq.~\ref{eq:mapping}, there are four possible terms since $x_i,x'_i \in \{0,1\}$. Each possibility corresponds to a one-qubit operator that can be expressed as the linear combination of the identity and at most two Pauli operators $\{\sx,\sy,\sz\}$ using the formulas
\begin{equation}\label{eq:op01}
|0\ra\la 1| = \half ( \sx + i\sy ) \equiv \hat \sigma^-, 
\end{equation}

\begin{equation}\label{eq:op10}
|1\ra\la 0| = \half ( \sx - i\sy ) \equiv \hat \sigma^+, 
\end{equation}

\begin{equation}\label{eq:op00}
|0\kb 0| = \half (\hat I + \sz),
\end{equation}

\begin{equation}\label{eq:op11}
|1\kb 1| = \half (\hat I - \sz).
\end{equation}

It has been shown previously that it is often advantageous to convert between encodings within the same Trotter step \cite{sawaya19_dlev}, though we do not consider encoding conversions in this work.

\subsection{Operators}
\label{sec:ops}

We study one- and two-particle operators for this work, for both bosonic and spin-$s$ degrees of freedom. Here we enumerate these operators and summarize properties that are relevant to resource requirements. We work in second quantization and define $\hat a^\dag$ and $\hat a$ respectively as the bosonic creation and annihilation operators. The bosonic position $\hat q = (\hat a^\dag + \hat a)/\sqrt 2$ and momentum $\hat p = i(\hat a^\dag - \hat a)/\sqrt 2$ operators are \textit{tridiagonal} matrices when represented in the number basis, as are spin-$s$ operators $S_x$ and $S_y$ when represented in the Z basis. The tridiagonality is relevant partly because the Gray code yields unity Hamming distance between integers $l$ and $l \pm 1$, which often leads to efficiency improvements over SB.

We explicitly show the sparsity patterns for $\hat q$ and $\hat q^2$ to aid in the interpretation of our numerical results:

\begin{equation}
\label{eq:q}
\begin{split}
\hat q = \frac{1}{\sqrt{2}}\begin{pmatrix}
0 & 1 & 0 & 0 & \dots \\
1 & 0 & \sqrt{2} & 0 & \dots \\
0 & \sqrt{2} & 0 & \sqrt{3} & \dots \\
0 & 0 & \sqrt{3} & 0 & \dots \\
\vdots & \vdots & \vdots & \vdots & \ddots
\end{pmatrix}
\end{split}
\end{equation}

\begin{equation}
\label{eq:qSq}
\hat q^2 = \half
\begin{pmatrix}
 1 & 0 & \sqrt{1\cdot2} & 0 &  \dots \\
 0 & 3 & 0 & \sqrt{2\cdot3} &  \dots \\
 \sqrt{1\cdot2} & 0 & 5 & 0 &  \dots \\
 0 & \sqrt{2\cdot3} & 0 & 7 & \dots \\
\vdots & \vdots & \vdots & \vdots  & \ddots
\end{pmatrix}.  
\end{equation}

Using the nomenclature introduced in \cite{sawaya19_dlev}, both the bosonic number operator $\hat n=\hat a^\dag a$ and the spin-$s$ $\hat S_z$ operator are diagonal binary-decomposable (DBD) and diagonal evenly spaced (DES), with the result that the SB encoding allows one to exactly implement their exponential form with only one-qubit gates when $d$ is a power of 2.

We include bosonic interaction operators $\hat n_i \hat n_j$, $\hat q_i \hat n_j$, and $\hat q_i \hat q_j$. We choose these three two-particle operators because they constitute all tensor product combinations of a diagonal and a banded tridiagonal matrix. Hence the results are applicable to other $d$-level systems with these operator characteristics---for example, results for $\hat n_i \hat n_j$ are similar to results for an operator $\Sint$ and $\hat q_i \hat n_j$ similar to results for $\hat S_x^{(i)} \hat S_z^{(j)}$. It is also notable that, though we do not show results in the paper, we found that resource counts for $\hat q_i \hat q_j$ and the bosonic interaction operator $\inttwo$ are nearly identical.

Finally, we note that although for many common problems $d<10$ tends to be sufficient, there are indeed cases where one requires $d$ to be 70 or greater \cite{sawaya19_vibr}.




\section{Bounds for SWAP gates}
\label{sec:ub}


\begin{figure}
    \centering
    \includegraphics[width=\linewidth]{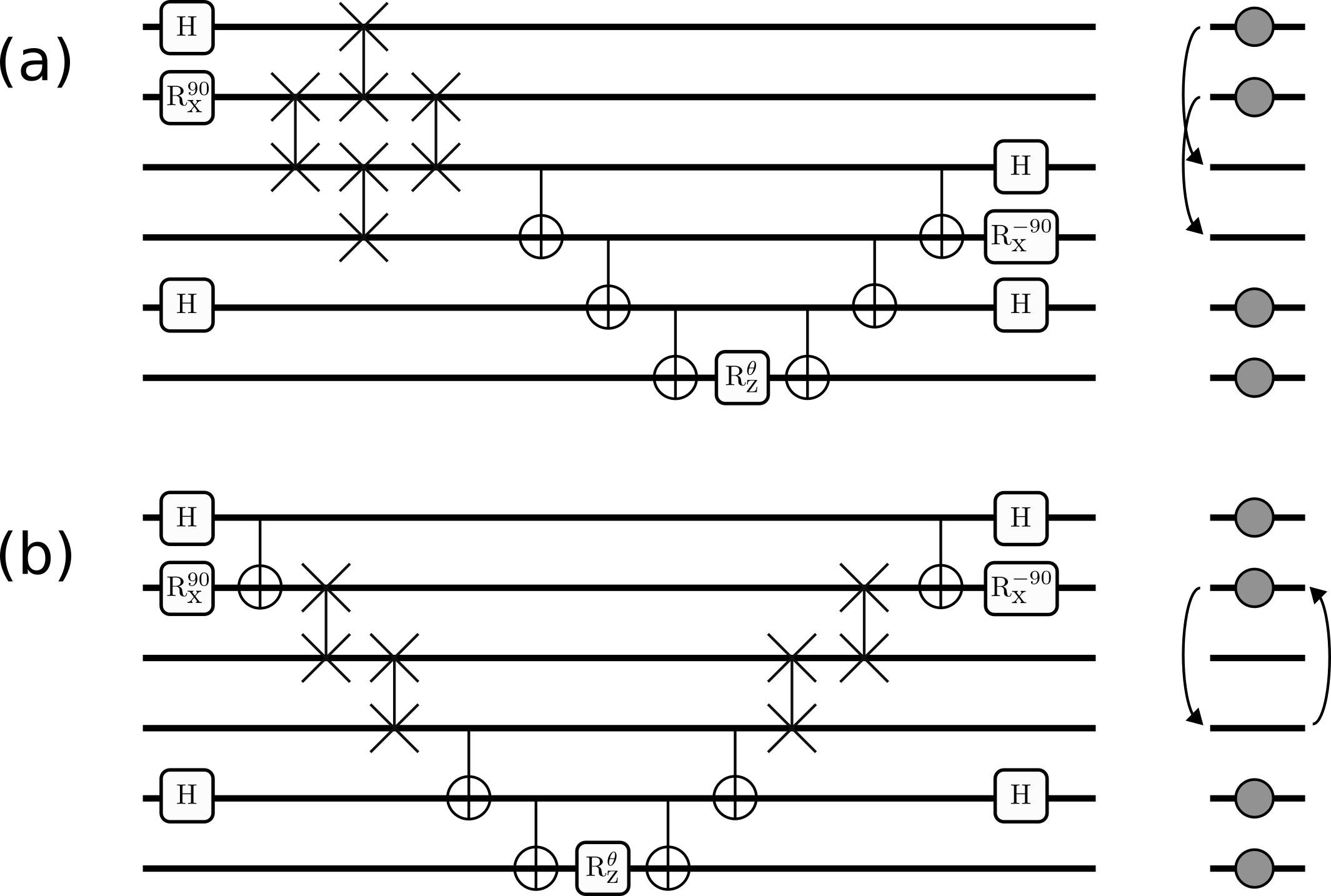}
    \caption{Two different reordering methods used in upper bound calculations for compact (Gray and SB) encodings. The circuits are implementing $\exp(-i \theta X_0\,Y_1\,X_4\,Z_5)$ on a 6-qubit device with linear connectivity and ordered initial placement. The 1-qubit gates at the beginning and end of the circuit are change of basis operations. 
    (a)~Cluster movement that reorders the qubits before completing the operation. 
    (b)~Shuttle movement that first moves a single program qubit down, then moves it back again to complete the operation. 
    The original placement is recovered at the end of (b), but not in (a).}
    \label{fig:shuttle-cluster}
\end{figure}



In this section we study upper bounds for the number of SWAP gates required. 
Our analytical results consider scheduling only on a 1D line. Analytical results are summarized in Table \ref{tab:compact_bounds} for SB/Gray encodings and Table \ref{tab:unary_bounds} for unary encodings.
Numerical results extend to higher-dimensional connectivity and are provided in Section~\ref{sec:numerical-results}.




{\renewcommand{\arraystretch}{1.3}
\begin{table}[]
\centering
\begin{tabular}{|l|l|l|}
\hline
\textbf{Density}         & \textbf{UB/LB}     & \textbf{Bound} (Compact) \\
\hline
Single term                     & UB          & $\half d (\log_2 d - h)$    \\
Dense [$d^2$]                     & UB          & $\frac{1}{8} (d \log_2^2 d - 2 d \log_2 d + 3d - 12)$   \\
Dense [$d^2$]                     & LB          & $\mathcal O(d /\sqrt{\log_2d}))$   \\
Dense 2-pcl                       & UB          & $\frac{1}{2} ( d^2 \log_2^2 d - d^2 \log_2 d + \frac{3}{4} d^2 - 3 )$   \\
Dense 2-pcl                       & LB          & $\mathcal O(d^2 / \sqrt{\log_2d})$   \\
\hline
\end{tabular}
\caption{\label{tab:compact_bounds} Summary of analytical upper/lower bounds (UB/LB) of SWAP gate counts, for compact (SB/Gray) encodings. Results were derived for the linear chain connectivity. The first column denotes the density of the $d$-level matrix operator; the first row refers to a single Hermitian term $\llpPLUSlpl$; $h$ is the Hamming distance; the last two rows refer to a product of two local $d$-level operators; `2-pcl' refers to a product of two local $d$-level operators.
}
\end{table}
}

{\renewcommand{\arraystretch}{1.6}
\begin{table}[]
\centering
\begin{tabular}{|l|l|l|}
\hline
\textbf{Density}         & \textbf{UB/LB}     & \textbf{Bound} (Unary) \\
\hline
Banded [$\mathcal O(d)$] & UB       & $\min( d^2 \frac{w-1}{4w} - d\frac{w-1}{4} , \frac{(d-w)w}{2} )$  \\
Banded 2-pcl                      & UB          & $d^2$  \\
Dense [$d^2$]                     & UB          & $\frac{d^2}{2} - \frac{3}{2} + 1$  \\
Dense 2-pcl                       & UB          & $d^2 + 2 N_{\textrm{UB,Unary}}^{w}$  \\
\hline
\end{tabular}
\caption{\label{tab:unary_bounds} Summary of analytical upper bounds of SWAP gate counts, for the unary encoding. Results were derived for the linear chain connectivity. The first column denotes the density of the $d$-level matrix operator; `2-pcl' refers to a product of two local $d$-level operators. 
}
\end{table}
}

\subsection{Compact operators}

Here we consider analytical upper bounds for compact codes, \textit{i.e.} those requiring $K=\ceil{\log_2 d}$ qubits per particle. For this analytical study, we assume that $d$ is a power of 2. $N_{\textrm{UB}}$ denotes the upper bound for the number of SWAP gates. We define $h$ as the Hamming distance between two bit strings. 


\textbf{Arbitrary Pauli string.} A single arbitrary Pauli string on $K$ qubits has length $p$, where the length is the number of non-identity Pauli operators. In order to implement the Trotter circuit, relevant qubits must be adjacent at some point in the calculation. In the worst case, the relevant qubits are split in two equally sized groups and separated as far as possible along the line. In calculating upper bounds, we consider two routes for determining a SWAP pattern for this worst case. As shown in  Figure \ref{fig:shuttle-cluster}, one may either consider a ``cluster move'' or a ``shuttle move.'' The former moves all the qubits next to each other, while the latter moves a qubit back and forth to complete the calculation. Asymptotically, this worst case requires SWAP counts of either
\begin{equation}
N_{\textrm{UB}}^{\textrm{Cl}}(p;K) = (K-p)\frac{p}{2}
\end{equation}
or
\begin{equation}
N_{\textrm{UB}}^{\textrm{Sh}}(p;K) = 2(K-p).
\end{equation}


In our first analysis, we assign $N_{\textrm{UB}}^{\textrm{Sh}}(p;K)$ to all Pauli strings of length $K$ and $p$ non-identity terms. This upper bound may be tightened in future work by considering how the Pauli operators are distributed among $K$ qubits.

\textbf{Single term.} Consider a real Hermitian operator with only two non-zero entries, $\llpPLUSlpl$. The mapping of one such term to a set of qubits leads to a sum of multiple Pauli strings. The distribution of Pauli string lengths is \cite{sawaya19_dlev}
%
\begin{equation}\label{eq:f_binom}
f(p; h,K) = \frac{1}{2} 2^{h} {K-h \choose p-h},
\end{equation}
where $f(p; h,K)$ is the number of length-$p$ Pauli strings and $h=h(\mathcal R(l),\mathcal R(l'))$ is the Hamming distance between the bit-string representations $\mathcal R$ of $l$ and $l'$. Note that $h\le p,K$.

The upper bound for the number of SWAP gates for one $\llpPLUSlpl$ term is thus
\begin{equation}\label{eq:cnt_llp}
\begin{split}
N_{\textrm{UB}}^{ll', h} &= \sum_{p=2}^K f(p; h,K)N_{\textrm{UB}}(p;K) \\
&= \sum_{p=2}^K 2^{h} {K-h \choose p-h} (K-p), \\ 
&= \half 2^K (K - h) \\
&= \half d (\log_2 d - h) 
\end{split}
\end{equation}




where $p=2$ is the first value of $p$ for which SWAP gates may be required.

A key result is that $N_{\textrm{UB}}^{ll', h}$ decreases as $h$ increases. This is the opposite of the analytical trend in CNOT counts \cite{sawaya19_dlev}. 
This is an intuitive result, since a higher $h$ leads to a higher density of Pauli terms, meaning that there are fewer ``gaps'' caused by local identity operators.

\textbf{All Pauli strings for $K$ qubits.} It is instructive to consider the highest possible bounds for $K$ qubits, for example in the worst case of encoding a fully dense matrix operator. At most, on $K$ qubits one may have all possible combinations of the four single-qubit (Pauli or identity) operators. Though there are at most $4^K$ unique Pauli strings, there are only $2^K$ combinations of $I$ and non-identity $\sigma \in \{X,Y,Z\}$. If one considers the latter number of strings, then the cluster movement is a more appropriate upper bound, as one may implement many Pauli matrix exponentials with one particular qubit ordering.





We consider all length-$m$ combinations of $\{I,\sigma\}$ leading up to $K$, $m \in \{3,4,...,K\}$. For a given $m$, the number of Pauli strings terminated with a non-identity is $2^m - 2^{m-1}=2^{m-1}$. This leads to bounds of

\begin{equation}
\begin{split}\label{eq:N_UB_compact_full_clus}
N_{\textrm{UB}}^{\textrm{All,}K} &= \sum_{m=3}^K \sum_i^{2^{m-1}} N_{\textrm{UB}}^{\textrm{Cl}}(p_i;m) \\
&= \sum_{m=3}^K \sum_i^{2^{m-1}} (m-p)p/2 \\
&< \sum_{m=3}^K 2^{m-1} m^2/8 \\
%
&= \frac{1}{8} (d \log_2^2 d - 2 d \log_2 d + 3d - 12)
\end{split}
\end{equation}


where using $p \leq m$ allowed us to write $(m-p)p \leq m^2/4$. 
This upper bound has lower complexity than $d$ times $N^{ll',h}_{UB}$ ($d \times \mathcal O(d \log_2 d)$), indicating that our original approach produces upper bounds that are too loose for full typical quantum operators of $\mathcal O(d)$ non-zero entries. However, note that a near-sighted scheduling algorithm might not be able to ``find'' the upper bound derived from cluster and/or shuttle movements, unless there are additional pre-compilation steps to order the Pauli strings in a favorable fashion.




For this particular case (all strings on $K$ qubits) we also derive lower bounds for SWAP counts. We consider only the number of SWAP gates required for the case of $p=K/2$, the $p$ with the highest number of distinct Pauli strings. Since any SWAP sequence used for this value of $p$ will result in many legitimate groupings for other values of $p$, limiting analysis to $p=K/2$ yields a lower bound.

There are $\binom{K}{K/2} \approx \frac{K!}{(K/2)!(K/2)!} \sim \mathcal O(2^K/\sqrt{K})$ unique strings of non-identity terms (\textit{i.e.} combinations using the two-member set $\{I,\sigma\}$), where in the last step we have used Stirling's approximation which is accurate even for very small $K$.

In any given placement on a line, there are exactly $K/2+1$ consecutive groups of length-$K/2$ strings (those starting from position 0, from 1, $\ldots$, and from $K/2$). A single SWAP yields \textit{at most} 2 new groups (neglecting the qubit order inside the group) and therefore even in the best case the SWAP counts needed are at least

\begin{equation}\label{eq:compact_lb}
N_{\textrm{LB}}^{\textrm{All,}K} \sim \mathcal O(2^K/\sqrt{K}).
\end{equation}

Hence we have relatively close lower ($\mathcal O(2^K/\sqrt{K})=\mathcal O(d /\sqrt{\log_2d})$) and upper ($\mathcal O(2^K K^2)=\mathcal O(d\log_2^2 d)$) bounds. 

\textbf{Two-particle operators.} Two-particle upper bounds for SWAP gate counts will display a similar trend with respect to the Hamming distance, because the same argument with respect to density of ``gaps'' applies. 
Here we wish to highlight the bounds for any coupled pair of one-particle operators, by considering the case of multiplying two $K$-qubit operators together where every possible Pauli string is present. Plugging $2K$ into equations \eqref{eq:N_UB_compact_full_clus} and \eqref{eq:compact_lb}, the worst case for complete $2K$ qubits is



\begin{equation}
\begin{split}
N_{\textrm{UB}}^{\textrm{All,2K}} =  \frac{1}{2} ( d^2 \log_2^2 d - d^2 \log_2 d + \frac{3}{4} d^2 - 3 ),
\end{split}
\end{equation}

and the lower bound is 

\begin{equation}
\begin{split}
N_{\textrm{LB}}^{\textrm{All,2K}} = \mathcal O(d^2 / \sqrt{\log_2d}).
\end{split}
\end{equation}

%
%


\subsection{Unary}

\textbf{One-particle operators.} Unlike the compact case, single-particle unary operators are linear combinations of Pauli strings with at most 2 non-identity factors. 
Exponentiating a single-particle unary operator consists of at most two-qubit operators. As a result, it is appropriate to take a different route than the compact case for estimating upper bounds.

\begin{figure}
    \centering
    \includegraphics[width=\linewidth]{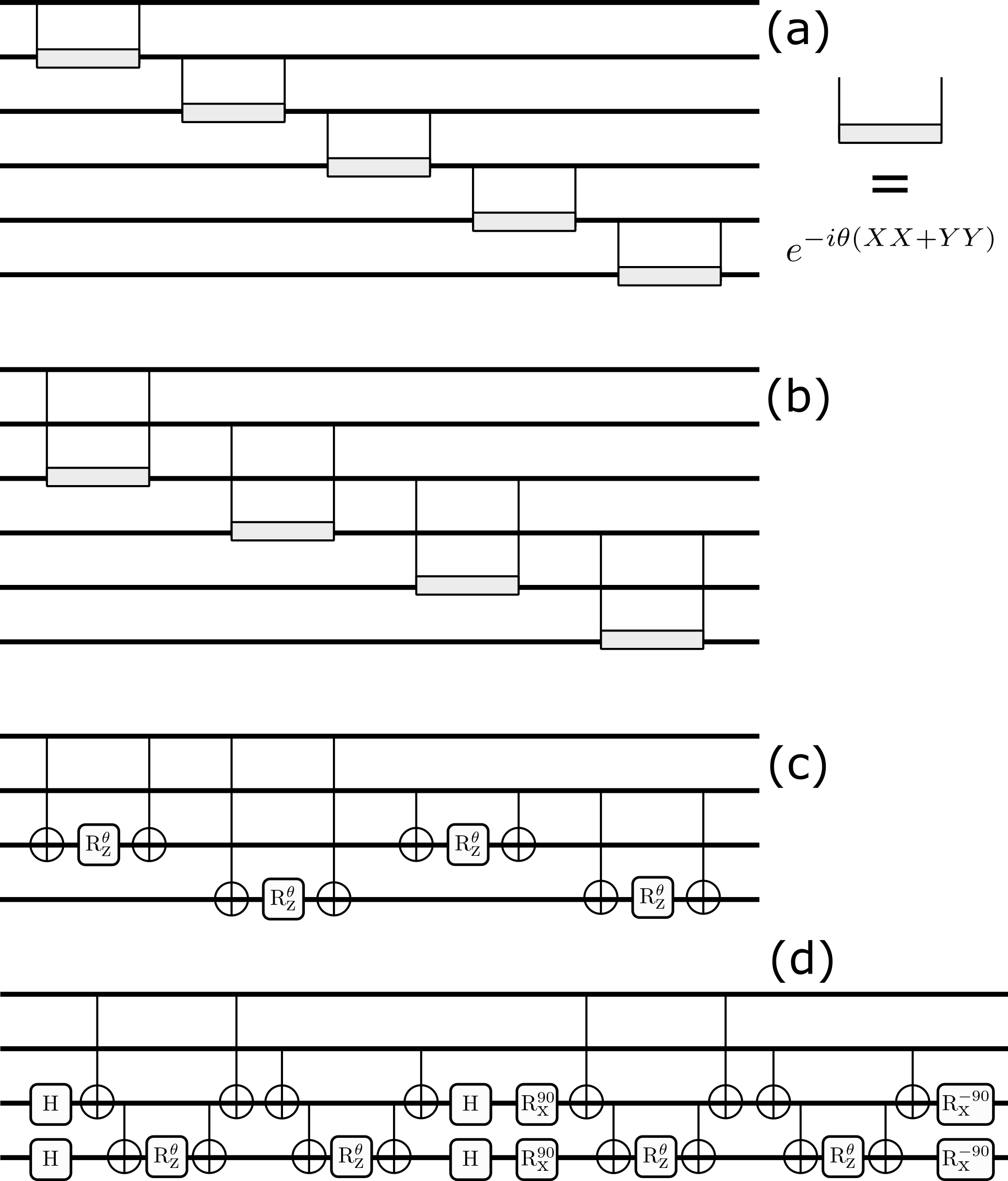}
    \caption{Trotter-Suzuki circuits for the unary encoding.
    (a)~One-particle banded operator with $w=1$ and $d=6$. 
    (b)~One-particle banded operator with $w=2$ and $d=6$.
    (c)~Two-particle interaction between two diagonal operators of 2-level systems. (d)~Two-particle interaction between a diagonal operator $Z_0 + Z_1$ and a $w=1$ operator $X_2 X_3 + Y_2 Y_3$.}
    \label{fig:unary_ordering}
\end{figure}

Consider an operator of $\mathcal O(d)$ terms, where terms occupy only a particular off-diagonal pair of bands, with the parameter $w=|l-l'|$ indicating the band including $\ket{l}\bra{l'}$. Common matrix operators tend to be ``banded'' in this way (\textit{e.g.} $w=1$ for $q$) or are sums of a small number of matrices with this characteristic (\textit{e.g.} $q^2$ is a sum of two matrices: one with only $w=0$ and one with only $w=2$). If $w$=1, then no SWAP gates are required when the qubits are in default ordering, as shown in Figure \ref{fig:unary_ordering}(a). If $w$=2, then at most two SWAP gates are required per term (Figure \ref{fig:unary_ordering}(b))---one per term to run the calculation, and one to return to the original position. 


The goal for a banded operator is to get qubits of appropriate CNOT indices to be adjacent, for example in the case of $w=2$ as in Figure \ref{fig:unary_ordering}(b). For a single-particle operator and a single band $w$, there are many orderings allowing one to implement all of the CNOT-Rz-CNOT motifs without SWAP gates. The adjacent CNOT indices naturally partition the qubits into $w$ distinct groups, each of which can be internally ordered so as to create all required adjacencies. One such ordering is the sequential placement of each group onto the linear connectivity graph (Table \ref{tab:inversions}).

To analyze the SWAP overhead of this configuration we use a well-known result of computer science: the number of required adjacent SWAP gates to convert one ordering to another is equal to the number of \textit{inversions} in the array \cite{kleinberg06_book}. Thus, given the ordering, the number of inversions required to re-order the circuit can be calculated directly.

\begin{table}[]
\centering
\begin{tabular}{|ccc|cc|}
\hline
 & $w=2$  & \# Inversions                & $w=3$   & \# Inversions  \\
\hline
 & 1    & 0                     & 1     & 0            \\
 & 3    & 0                     & 4     & 0            \\
 & 5    & 0                     & 7     & 0            \\
 & 7    & 0                     & 10    & 0            \\
 & 9    & 0                     & 2     & 3            \\
 & 11   & 0                     & 5     & 2            \\
 & 2    & 5                     & 8     & 1            \\
 & 4    & 4                     & 11    & 0            \\
 & 6    & 3                     & 3     & 6            \\
 & 8    & 2                     & 6     & 4            \\
 & 10   & 1                     & 9     & 2            \\
 & 12   & 0                     & 12    & 0            \\
 \hline
 & \textbf{Sum}: & \textbf{15}                    & \textbf{Sum}:  & \textbf{18} \\
 \hline
\end{tabular}
\caption{\label{tab:inversions} Determining upper bounds for SWAP gates needed to Trotterize banded sparse one-particle operators, for 1D linear connectivity. Each left-hand column provides a qubit index ordering for which one does not need SWAP gates to implement a Suzuki-Trotter step, where $w$ denotes which off-diagonal band of the matrix operator is non-zero ($w=2$ is shown in Figure \ref{fig:unary_ordering}(b)). w=1 is depicted in equation \eqref{eq:q} and w=0,2 are depicted in equation \eqref{eq:qSq}. Each right-hand column gives the number of ``inversions'' relative to the default qubit ordering of $\{0,1,2,\dots\}$.
}
\end{table}

Table \ref{tab:inversions} counts inversions for different values of $w$. From inspection, the upper bound for the number of SWAP gates will be
\begin{equation}
N_{\textrm{UB,Unary}}^{w,\textrm{small }d} = \sum_{g=1}^{w-1} \bigg(g \sum_{j=1}^{\ceil{d/w}-1} j \bigg)
\end{equation}


which leads to

\begin{equation}\label{eq:inversion_ub}
N_{\textrm{UB,Unary}}^{w,\textrm{small }d} = d^2 \frac{w-1}{4w} - d\frac{w-1}{4}
\end{equation}

It is notable, especially when considering low $d$ values, that as $w$ increases the second (linear) coefficient increases faster than the first (quadratic) coefficient.


A simpler route leads to linear scaling. One may instead consider every relevant pair of qubits, SWAP to make them adjacent, and then SWAP back to the original ordering once the two-qubit exponentiation has been performed pair, before moving on to the next. This leads us to


\begin{equation}\label{eq:unary_linear_ub}
N_{\textrm{UB,Unary}}^{w,\textrm{large }d} = 2(d-w)w
\end{equation}

Expression \eqref{eq:unary_linear_ub} does not lead to lower SWAP counts than \eqref{eq:inversion_ub} until at least $d\ge30$.

\textbf{Dense one-particle operator.} We consider the case of a fully dense matrix in the unary encoding, in which all pairs of $d$ qubits must be adjacent at some point in the calculation, in order to exponentiate all terms $X_iX_j+Y_iY_j$. In such a case, one can implement a linear-depth SWAP network based on previous work in \cite{holmes18_connect} that considered SWAP gates required for the quantum Fourier transform (QFT). 
The SWAP pattern for QFT leads to a SWAP count upper bound of
\begin{equation}
    N_{\textrm{UB,Unary}}^{\textrm{Dense,1pcl}} = \frac{d^2}{2} - \frac{3}{2} + 1.
\end{equation}



These SWAP gates alone may be implemented in depth $2d-3$ \cite{holmes18_connect}.

\textbf{Two-particle operators.} Because a one-particle operator will include at most two-qubit terms, a two-particle operator (as it is built from products of one-particle operators) will include at most four-qubit terms. A few cases are instructive to consider. A product of two diagonal operators has structure $a_0 Z_0Z_{d} + a_1 Z_0Z_{d+1} + \cdots$, which yields at most $d^2$ two-qubit terms. The product of a diagonal and a banded off-diagonal ($w>0$) operator has structure $a_0 Z_0(X_{d}X_{d+w} + Y_{d}Y_{d+w}) + a_1 Z_0(X_{d+1}X_{d+w+1} + Y_{d+1}Y_{d+w+1}) \cdots$, leaving at most $2(d-w)d=2(d^2-dw)$ three-qubit terms. And the product of two banded operators (with equal $w>0$) has structure $(X_{0}X_{w} + Y_{0}Y_{w})(X_{d}X_{d+w} + Y_{d}Y_{d+w}) + \cdots$, leading to at most $(2(d-w))^2=4(d^2-2dw-w^2)$ four-qubit terms. Examples of Trotterized two-particle operators are given in Figures \ref{fig:unary_ordering}(c) and \ref{fig:unary_ordering}(d). Because of the $X_iX_j+Y_iY_j$ motif, multiple Pauli strings can be exponentiated with the same qubit placements, a fact we take advantage of in our upper bound calculations.

Unlike in the single-banded one-particle unary case, one cannot execute all of the exponentials of the two-particle case with a single ordering, as there are $\mathcal O(d^2)$ terms. For the diagonal-diagonal interaction operator (\textit{e.g.} $\ninj$), we note that the interactions form a complete bipartite graph with $d$ nodes in each partition. Beginning with ordering $\{0,1,\cdots,d,d+1,\cdots,2d-1\}$, one may move the qubits of the bottom particle up until one reaches ordering $\{d,d+1,\cdots,2d-1,0,1,\cdots\}$. On the way to the final ordering, all relevant pairs of qubits are adjacent at some point in the procedure. This requires $d^2$ SWAP operations.

Next we consider the product of two banded one-particle operators (\textit{e.g.} $\qiqj$), both with $w=1$. 
In this case, the same SWAP procedure from the diagonal-diagonal case may be used. 
Note that both terms like $(X_1X_{2} + Y_1Y_{2})(X_{10}X_{11} + Y_{10}Y_{11})$ and $(X_1X_{2} + Y_1Y_{2})(X_{11}X_{12} + Y_{11}Y_{12})$ need to be considered. As the second particle's qubits are moved upwards, all relevant combinations of four qubits can be made adjacent at some point, though in an order that may be different from how the Hamiltonian term was originally written down. 
For instance, if $\{1,2,10,11\}$ is an ordering at one point in the swap network, then a few SWAPs later an ordering $\{1,11,2,12\}$ will naturally appear. This does not affect the calculation, as the only requirement for exponentiating a Pauli string is that the relevant qubits are adjacent.

We note this upper bound for SWAP operations required:

\begin{equation}
N_{\textrm{UB,Unary}}^{\textrm{2pcl,diag-diag}} = N_{\textrm{UB,Unary}}^{\textrm{2pcl,}w=1} = d^2.
\end{equation}

Methods used in previous work \cite{holmes18_connect} can be used to show that this sequence of SWAPS can be performed in depth $2d-1$.

In the case of a product of two banded operators with $w>1$, one may calculate an upper bound for the SWAP count by considering the reordering of each particle's qubits to the order $\{0,w,2w,\cdots,1,w+1,2w+1,\cdots\}$. This leaves us with the expression

\begin{equation}
N_{\textrm{UB,Unary}}^{\textrm{2pcl,}w>0} = d^2 + 2 N_{\textrm{UB,Unary}}^{w}
\end{equation}

where either equation \eqref{eq:inversion_ub} or \eqref{eq:unary_linear_ub} may be used for $N_{\textrm{UB,Unary}}^{w}$ and the factor of 2 in the second expression is due to the presence of two particles.



\subsection{Comparisons between unary and compact}

Though the upper bounds for compact and unary were calculated using different routes, both of which led to relatively loose bounds, it may be somewhat useful to make tentative comparisons between them. %

Equation \eqref{eq:inversion_ub} scales worse than the absolute upper bound for the compact case calculated in \eqref{eq:N_UB_compact_full_clus}. This is partly because more qubits are present in the unary case, often leading to more distance that must be travelled. However, one often is not interested in increasing $d$ asymptotically, as most physics applications involve building up a system of particles with bounded $d$. The fact that $\half d \log_2^2 d$ (equation \eqref{eq:N_UB_compact_full_clus}) is larger than $d^2 \frac{w-1}{4w} - d\frac{w-1}{4}$ (equation \eqref{eq:inversion_ub}) for small $d$ suggests that unary may often have fewer SWAP gates.

Note that it was previously shown that unary usually requires fewer CNOT gates than compact codes for Trotterization. Hence this comparison of SWAP gates is relevant because it suggests that unary, at least for low-$w$ operators that are common in quantum simulation, will often have superior overall gate counts for both all-to-all connectivity (which consider only CNOT gates) and linear connectivity (which consider both CNOT and SWAP counts). As we will show, the numerical results of this work bear out some of these trends.

Another relevant fact is that Pauli strings of the unary encoding always have length $p \leq 4$, assuming one deals with two-particle operators at the most. One would expect this to affect circuit depth, as it means that more Pauli exponentials can be implemented in parallel in the unary as compared to the compact case.

\section{Numerical Methods}
\label{sec:methods}

\begin{figure}
    \centering
    \includegraphics[width=0.4\linewidth]{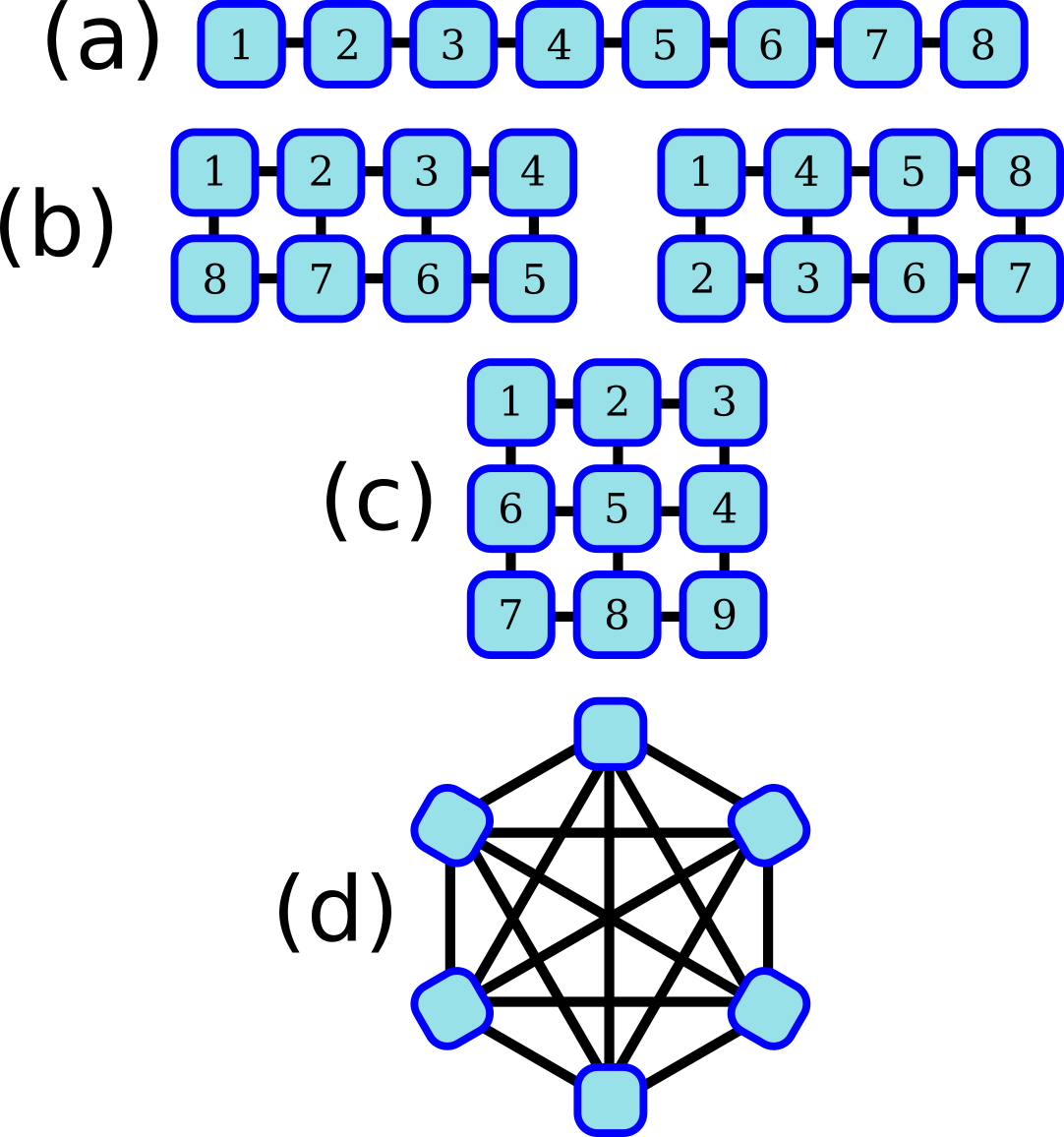}
    \caption{Qubit connectivities. From top to bottom: (a) Linear chain, (b) ladder, (c) square grid, and (d) all-to-all. In (b), both ``horizontal snake'' and ``vertical snake'' placements are shown.}
    \label{fig:connect}
\end{figure}

We consider three connectivities in this work: linear, ladder, and square grid. 
These are shown in Figure \ref{fig:connect} together with the all-to-all case.

To determine the series of SWAP gates required in the routing process, we adopt the scheduler of quantum circuits described in \cite{guerreschi18_scheduling,guerreschi19_scheduler}. The inputs to the scheduling algorithm are the quantum circuit, the hardware's connectivity, and the initial placement of program qubits on the physical qubits. When a two-qubit gate between unconnected qubits needs to be executed, the scheduler considers all possible SWAP gates available in the architecture (i.e. one SWAP per edge of the connectivity graph) and adds those that reduce the distance between the program qubits involved in the two-qubit gates. When multiple SWAPs have the same utility, the scheduler chooses one of them according to a greedy stochastic policy. 
Due to this stochasticity, for each distinct input of the scheduler, we generate $>1000$ 
stochastic schedules and report the one with the minimum number of SWAP gates.

Our initial qubit placement for the linear connectivity was the standard ordering $0,1,2,\dots$. For the ladder, we considered the ``vertical snake'' and ``horizontal snake'' placements explicitly visualized in Figure~\ref{fig:connect}, and reported the lowest SWAP count that the scheduler was able to find. When the square grid is implemented, the length of the sides are $\ceil{\sqrt{N_q}}$, we also use the snake placement. These initial positions are arbitrary, since an operation will most often be run after previous operations that have already re-ordered the qubits. However, choosing a consistent starting placement allows for direct comparisons between different encodings, and trends with respect to operator characteristics (such as sparsity structure) will often be applicable regardless of starting placement. 

In our numerical results, we do not consider the depth of the circuits. This is because optimizing depth requires an entirely new set of considerations, namely the order in which the single exponentials $\exp(-i\hat h_j \tau/\eta)$ are implemented. This leads to a rich set of additional optimization considerations that we leave for future work. We note that one would expect the SWAP counts to be correlated with the circuit depth and its minimization a good proxy for overall circuit fidelity \cite{Sivarajah2020a}.

\section{Numerical Results}
\label{sec:numerical-results}

The scheduling process took between a few seconds to almost one hour on a dual-socket Xeon Platinum 8280 (28 cores Cascade Lake per socket at 1.8GHz AVX base frequency, turbo disabled, 192 GB, dual-rail OPA) depending on the input.

\subsection{One particle with linear connectivity}
\begin{figure}
    \centering
    \includegraphics[width=\linewidth,trim=0 100 0 100, clip]{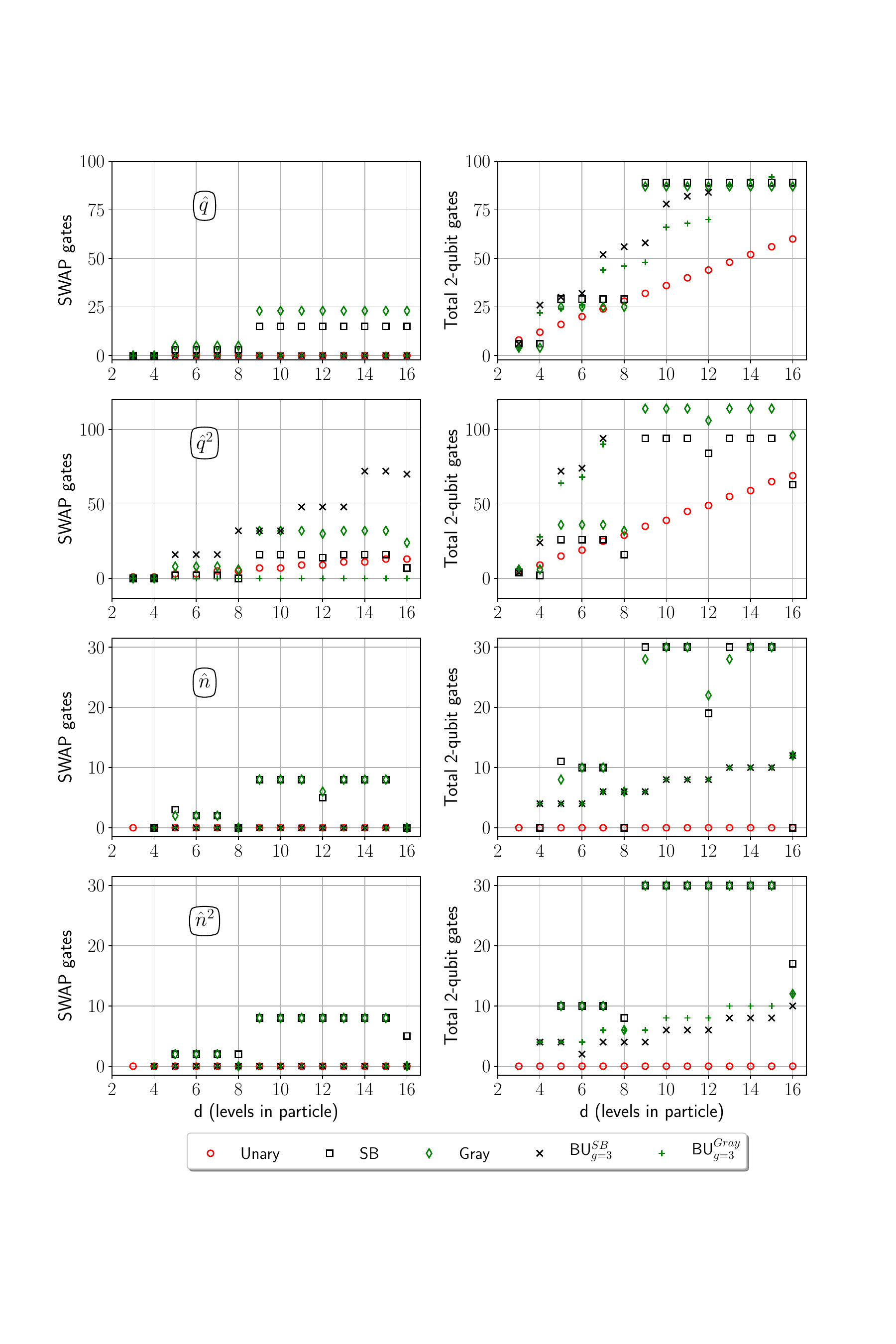}
    \caption{SWAP gate counts and total two-qubit gate counts for a Suzuki-Trotter step of single-particle bosonic operators with increasing truncation $d$, using \textit{linear} hardware connectivity. Some data points are cut off in order to show low-$d$ orderings more clearly.
    }
    \label{fig:1pc-cnts}
\end{figure}

Figure \ref{fig:1pc-cnts} shows gate counts for one Suzuki-Trotter step of single-particle bosonic operators. The left column shows SWAP gate counts while the right column counts total two-qubit gates (SWAP and CNOT). Even though the Gray code requires more SWAP gates than SB (consistent with analytical arguments), for the tridiagonal $\hat q$ it usually requires slightly fewer total two-qubit gates. For $\hat q^2$, the larger number of SWAP gates leaves Gray inferior for all simulated $d$ values, when using the greedy scheduler.

For $\hat q$, the unary code is improved relative to the all-to-all connectivity case, as it does not require any SWAP gates. Even in the case of $\hat q^2$, fewer SWAP gates provides it with a larger advantage over compact codes than before. For diagonal operators $\hat n$ and $\hat n^2$, the unary always requires only single-qubit gates, and hence no SWAP gates are required either. Gray and SB yield similar results, because the extra SWAP gates required for Gray mostly cancel out its previous advantage in all-to-all connectivity.

The visibly favorable decrease in operations counts when $\log_2 d$ is an integer has been discussed previously \cite{sawaya19_dlev}, inclduing the fact that the bosonic $\hat n$ is always most efficient with SB, requiring only one-qubit gates.

When dealing specifically with bosonic degrees of freedom, this implies that increasing the truncation often is helpful. For instance, if a problem requires a truncation of $d=5$ or greater, then it is appropriate to compare unary's $d=5$ gate counts to the lowest compact (SB or Gray) gates counts in the range of $d=5$--8, as all of the latter require the same number of qubits. In all plotted results, we show the values that result for truncating the given matrix at $d$ (\textit{e.g.} if Gray's gate counts for $d=8$ are less than for $d=5$, we still plot the original $d=5$ value). This choice was made in order to show behavior for general matrix-to-qubit mappings with these sparsity structures, as some problems types (like spin-$s$ or classical combinatorial problems) do not allow one to increase the truncation.

\begin{figure}
    \centering
    \includegraphics[width=\linewidth,trim=0 0 0 50, clip]{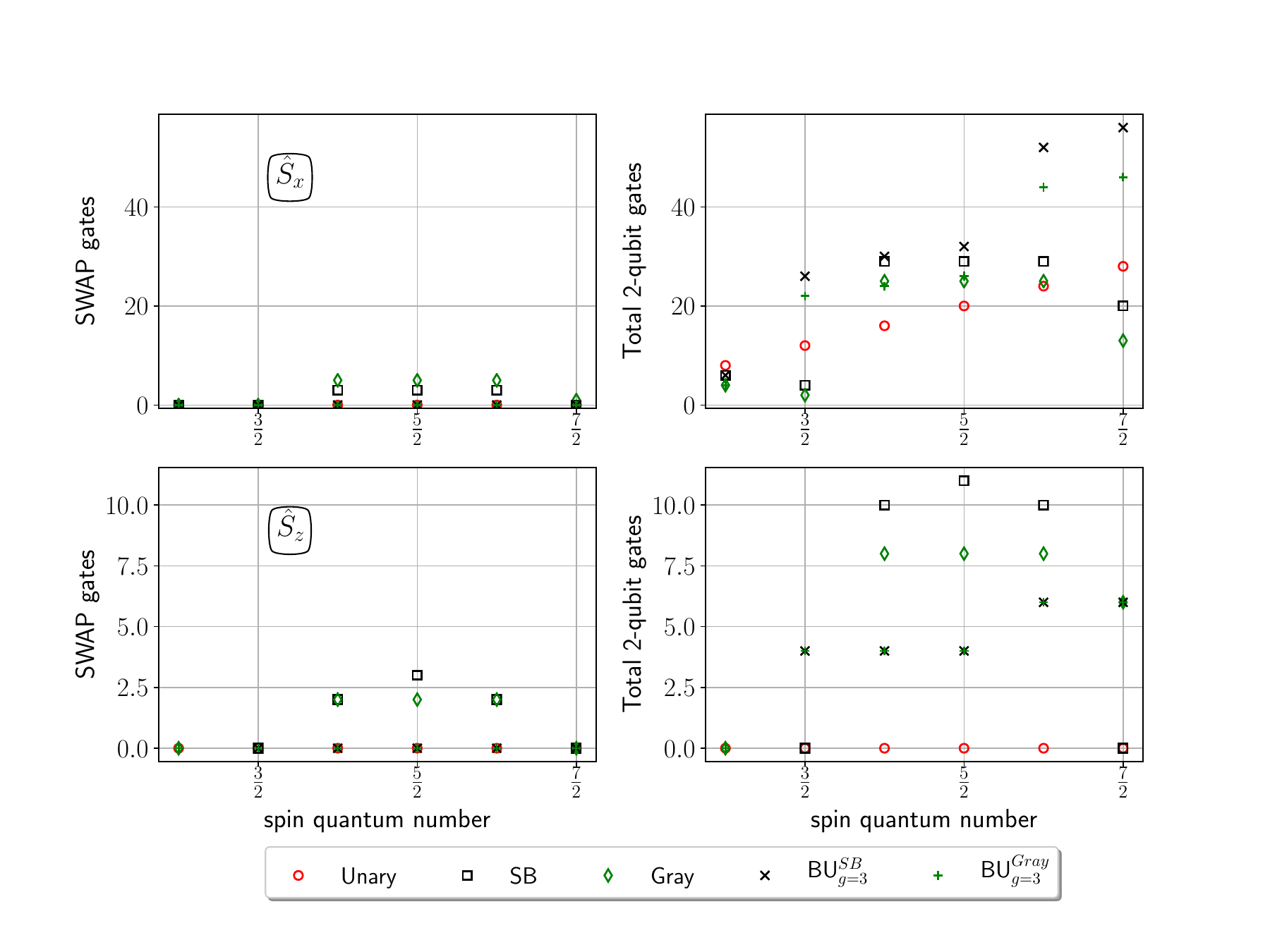}
    \caption{SWAP gate counts and total two-qubit gate counts for a Suzuki-Trotter step of single-particle spin-$s$ particles, using \textit{linear} hardware connectivity. 
    }
    \label{fig:sxsz-cnts}
\end{figure}

\begin{figure}
    \centering
    \includegraphics[width=\linewidth,trim=0 0 0 100, clip]{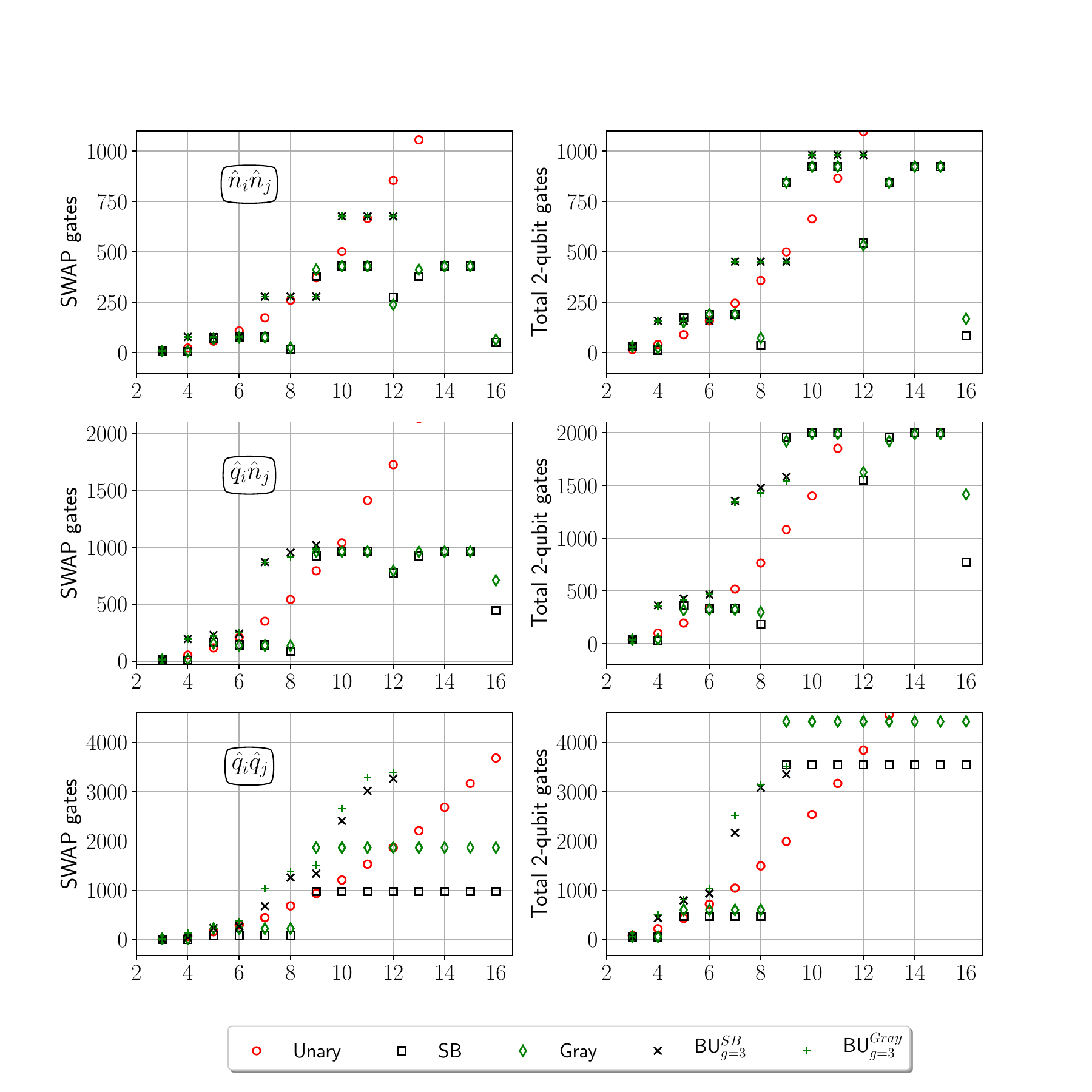}
    \caption{SWAP gate counts and total two-qubit gate counts for a Suzuki-Trotter step of adjacent two-particle bosonic operators, using \textit{linear} hardware connectivity. Some data points are cut off in order to show low-$d$ orderings more clearly. 
    }
    \label{fig:2pcl-cnts}
\end{figure}

\begin{figure}
    \centering
    \includegraphics[width=\linewidth]{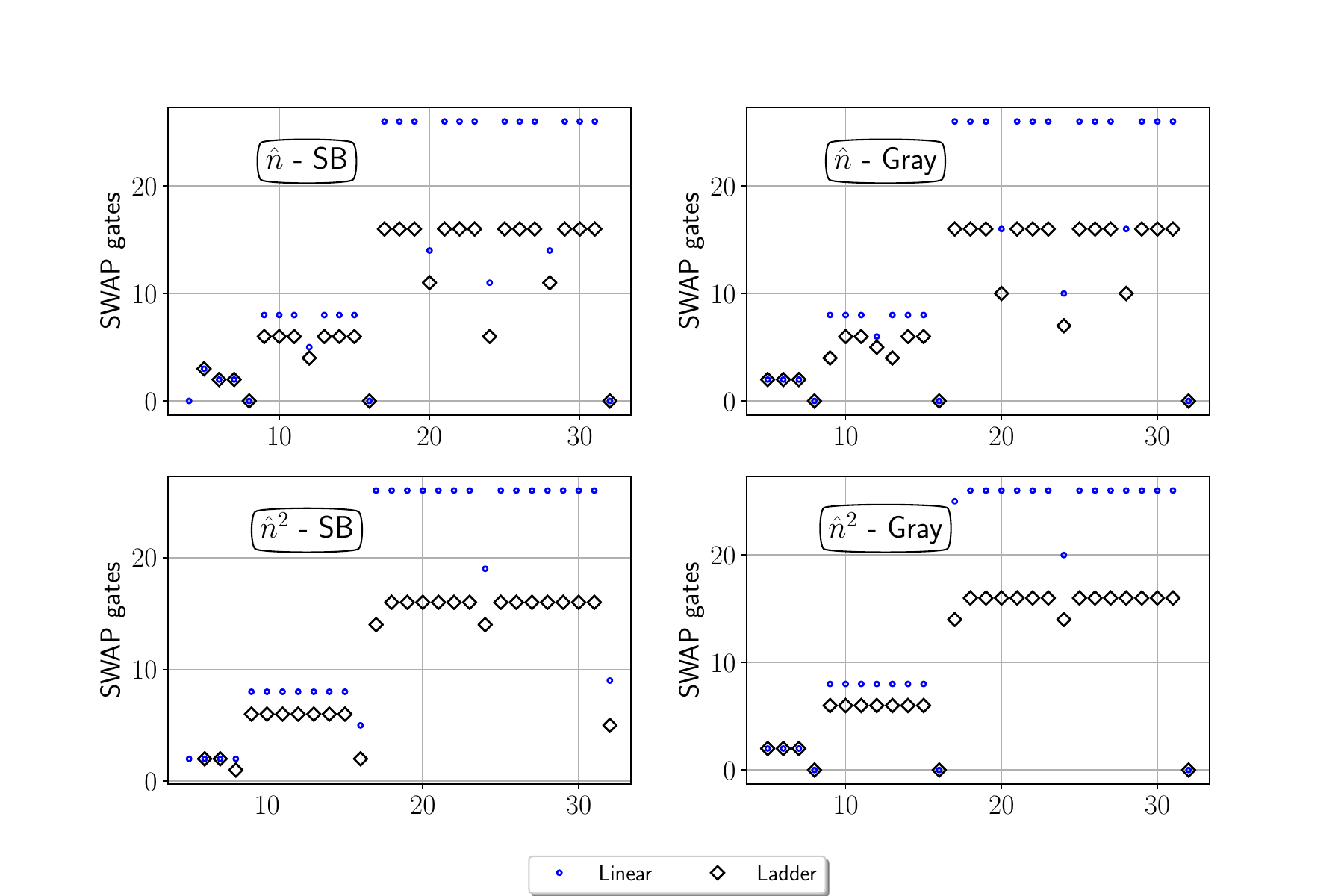}
    \caption{SWAP gate counts for a Suzuki-Trotter step of the bosonic $\hat n$ and $\hat n^2$ operators, for different encodings, hardware connectivities, and initial qubit placements.}
    \label{fig:plc-n-nsq}
\end{figure}

\begin{figure}
    \centering
    \includegraphics[width=\linewidth]{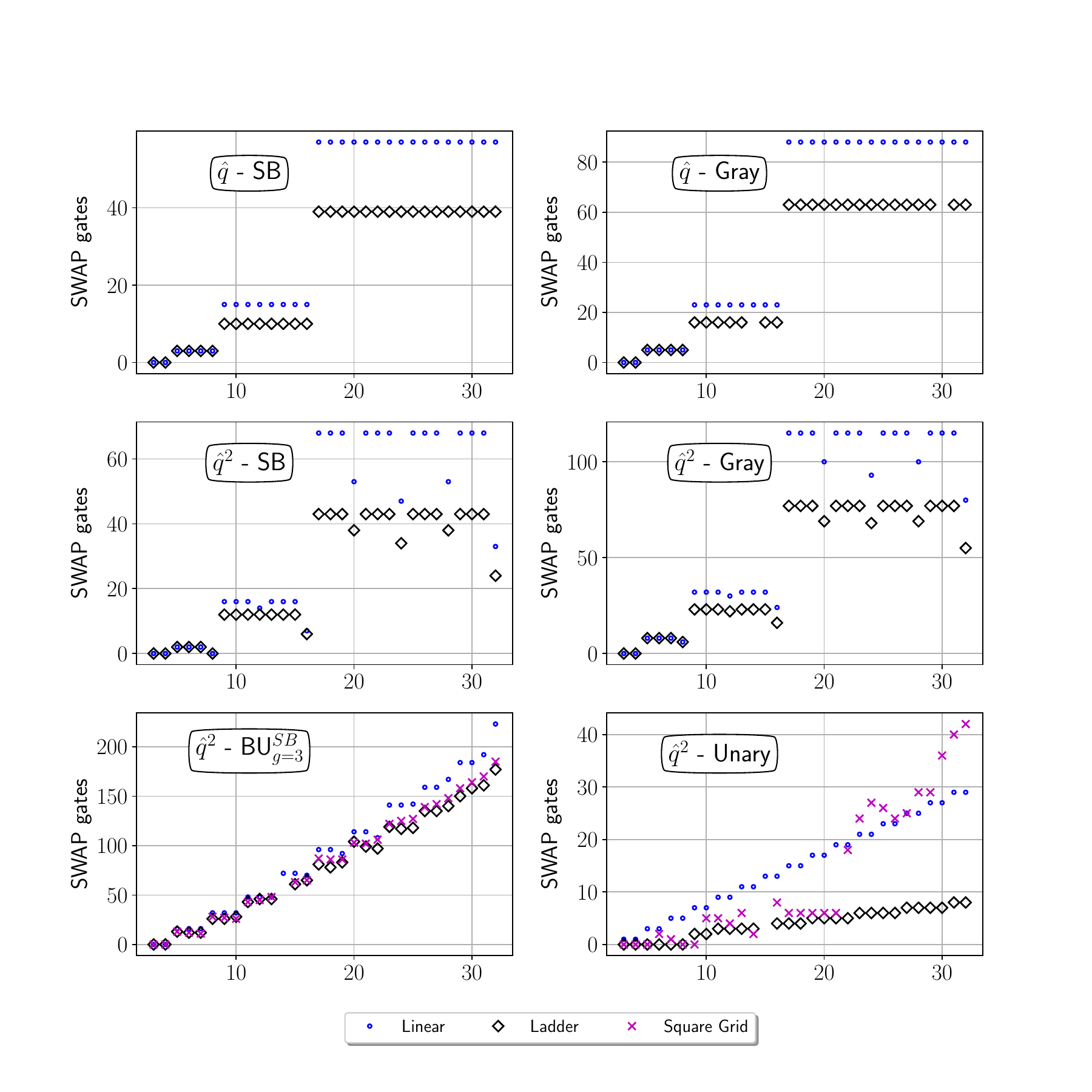}
    \caption{SWAP gate counts for a Suzuki-Trotter step of the bosonic $\hat q$ and $\hat q^2$ operators, using compact encodings, for different hardware connectivities and initial qubit placements. Note that different encodings have vastly different qubit counts. 
    }
    \label{fig:plc-q-qsq}
\end{figure}

\begin{figure}
    \centering
    \includegraphics[width=\linewidth]{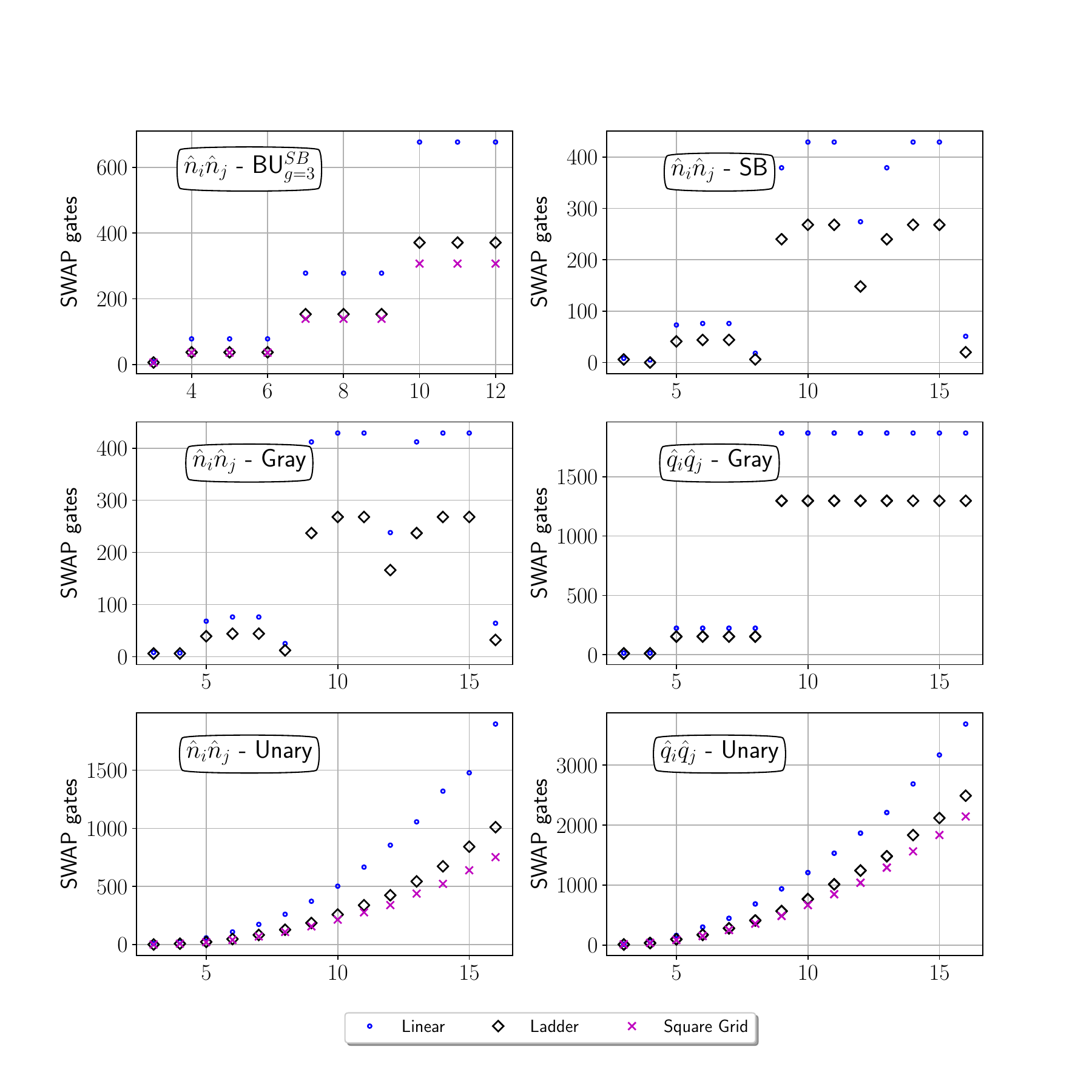}
    \caption{SWAP gate counts for a Suzuki-Trotter step of the bosonic $\hat n$ and $\hat n^2$ operators, for different encodings, hardware connectivities, and initial qubit placements. Note that different encodings have vastly different qubit counts. 
    }
    \label{fig:plc-2pcl}
\end{figure}

Figure \ref{fig:sxsz-cnts} shows results for a single-particle Suzuki-Trotter step of $\Sx$ and $\Sz$. In this case, one may not increase the truncation, as $d$ is determined by the inherent properties of the particle (its spin $s$). There are no particularly strong trends, except that the Gray code tends to be slightly favorable despite its higher SWAP gate counts. We omit plots for two-particle spin-$s$ operators, since the trends are similar to bosonic interaction operators (\textit{e.g.} $\SxSx$ $\sim$ $\qiqj$).

Note that there is a limited set of instances where block unary \textit{may} prove useful for specific hardware. These are the cases in which a BU code requires fewer qubits than the unary \textit{and} fewer operations than the compact codes (\textit{e.g.} $d=9$ through 12 for $\hat q$). These limited use cases will be hardware-dependent, and are best thought of in terms of the hardware budgets as discussed previously \cite{sawaya19_dlev}.

\subsection{Two particles with Linear connectivity}

Figure \ref{fig:2pcl-cnts} shows SWAP gate counts and total two-qubit gate counts for Suzuki-Trotter circuits of $\hat n_i \hat n_j$, $\hat q_i \hat n_j$, and $\hat q_i \hat q_j$. Though it is not shown, we note that numerical results for $\inttwo$ were nearly identical to those for $\hat q_i \hat q_j$. The total number of operations increases as the number of non-diagonal operators (\textit{i.e.} $\hat q$) increases from 0 to 2, because off-diagonal terms lead to more complex Pauli Hamiltonians.

In the unary code, the empirical trend for the SWAP gate counts is now super-linear, consistent with the analytics, and consistent with the fact that there are $\mathcal O(d^2)$ Pauli strings instead of $\mathcal O(d)$ in the single-particle unary case.

When one allows for increasing the bosonic truncation for compact codes (as discussed previously), unary-encoded $\ninj$ and $\qinj$ do not appear to be advantageous for any $d$. Considering $\qinj$ for example, unary with $d=9$ requires more gates than SB at $d=16$.

On the other hand, for $\qiqj$, the optimal encoding is $d$-dependent, with unary often requiring the fewest total two-qubit gates. Finally, as in the one-particle case, there are very limited cases (such as $d=9$ for all shown two-particle operators) where BU may be occasionally advantageous for near-term hardware.

\subsection{Varying hardware connectivity}

We now compare the scheduler's results across the three hardware connectivities. 
Each plot in this section studies one encoding-operator pair while varying $d$. Only SWAP gate counts are compared. 
We do not include square grid results for compact encodings, as the ladder and square grid layouts are either identical or very similar for small qubit counts.

Figure \ref{fig:plc-n-nsq} shows results for compact representations of $\hat n$ and $\hat n^2$, for which all data points use at most $N_q=5$ qubits. Unary is excluded because it yields zero SWAP gates. A substantial improvement is shown due to the ladder connectivity. 

Figure \ref{fig:plc-q-qsq} shows results for a selection of encodings of $\hat q$ and $\hat q^2$. Ladder grid improvements for the compact codes are substantial and similar in magnitude to the diagonal cases. The unary case for $q^2$ shows substantial improvements when switching from linear to ladder, though the switch to square grid is either not advantageous or is detrimental. The square grid results become especially poor around the point that a 5-by-5 qubit grid is needed. This is intriguing, because a linear schedule can always be mapped to a snaked qubit placement on a grid, implying that linear should never be better than a higher connectivity. We conclude that the near-term decisions of the scheduler are made to the detriment of future gates in the quantum circuit since no look-ahead mechanism is present. 

Figure \ref{fig:plc-2pcl} gives SWAP gate counts for a selection of two-particle operators. $\qinj$ is excluded because results are very similar to $\ninj$ and $\qinj$. Again, we see diminishing returns when increasing connectivity.

\section{Discussion \& Outlook}
\label{sec:concl}

Using analytical upper bound calculations as well as a numerical quantum circuit scheduler, we have studied the connectivity-dependent two-qubit operation counts required to approximate the exponentials of common bosonic and spin-$s$ operators.

Most scaling trends derived in the upper bounds calculations were observed in the numerical data. Importantly, the Gray code indeed required more SWAP gates than the standard binary code, due to the former's lower-length Pauli strings. The unary code requires more SWAP gates the further its matrix elements are from the diagonal. The optimal encoding for two-particle operators is closely dependent on $d$, though compact encodings tends to be superior.

Compared to the previously studied all-to-all connectivity, the low-$d$ advantage of the unary code is slightly increased. As was the case before, block unary code is optimal only for a narrow set of operator-$d$ pairs. Interestingly, the Gray code is not as advantageous as previously predicted, often in fact showing a higher total two-qubit gate counts than standard binary. This is surprising, because the Gray code usually produces shorter Pauli string lengths. It may be that a scheduler that looks beyond the local optimum would find a more beneficial schedule for Gray-encoded circuits. We also note that these simpler Pauli Hamiltonians still suggest that the Gray code will be advantageous for algorithms like the variational quantum eigensolver, as simpler Hamiltonians require fewer measurements.

We tested three connectivities: linear, ladder, and square grid. In most cases, an increase in dimensionality lowers the number of required SWAP gates.
Notably, changing from linear to ladder leads yielded a much larger percent improvement than changing from ladder to grid. Additionally, in some cases the scheduler's results for the ladder are superior to the square grid. This is partially attributable to the scheduler considering a limited number of upcoming gates.

Fruitful future work may include tightening upper bounds by considering the fine-grained distribution of single-qubit Pauli operators. Additionally, the scheduler may be improved by taking more future gates into account, or by actively constructing the Trotterized quantum circuit using the qubit Hamiltonian as a starting point. Further, it will be important to study circuit depth for different hardware connectivities, which would require more detailed consideration of the order of the Suzuki-Trotter product. Finally, these results may be applicable for scheduling more general quantum algorithms (unrelated to physics) for which matrix exponentiation is a subroutine \cite{gharibian15_qhamcompl}. 

Notably, there are instances for which each one of the studied encodings is the preferable choice. This fact highlights the need to consider multiple encoding types when preparing for Hamiltonian simulation of $d$-level particles, and multiple hardware connectivities when considering hardware design.

\bibliographystyle{ieeetr}
\bibliography{references}


\end{document}